\documentclass{egpubl}
\usepackage{eurovis2024}

\EuroVisShort  

\usepackage[T1]{fontenc}
\usepackage{dfadobe}  

\usepackage{cite}  
\BibtexOrBiblatex
\electronicVersion
\PrintedOrElectronic
\ifpdf \usepackage[pdftex]{graphicx} \pdfcompresslevel=9
\else \usepackage[dvips]{graphicx} \fi

\usepackage{egweblnk}

\title{Interaction Techniques for Exploratory\\~Data Visualization on Mobile Devices}

\author[Snyder~\etal]
{
    \parbox{\textwidth}{
        \centering %
        Luke S. Snyder$^{1}$, %
        Ryan A. Rossi$^{2}$, %
        Eunyee Koh$^{2}$, %
        Jeffrey Heer$^{1}$, %
        and Jane Hoffswell$^{2}$%
    }
    \\
    {
        \parbox{\textwidth}{
            \centering %
            $^1$University of Washington, USA
            $^2$Adobe Research, USA\\%
        }
    }\vspace{-110px}
}

\usepackage{xargs}      
\usepackage{soul}       
\usepackage{color}      
\usepackage{xspace}     
\usepackage{xpunctuate} 
\usepackage{booktabs}
\usepackage{amsmath}
\usepackage{multirow}

\newcommand{\ie}{{i.e.,}\xspace}
\newcommand{\eg}{{e.g.,}\xspace}
\newcommand{\etal}{{et~al\xperiod}\xspace}

\newcommand{\bpstart}[1]{\vspace{0px}\noindent\textbf{#1}}

\newcommand{\myquote}[1]{\emph{``#1''}}                         

\definecolor{lightpink}{RGB}{237,157,202}
\definecolor{lightred}{RGB}{210,121,121}
\definecolor{lightorange}{RGB}{230,170,50}
\definecolor{lightgold}{RGB}{210,194,121}
\definecolor{lightgreen}{RGB}{121,210,121}
\definecolor{lightaqua}{RGB}{121,206,210}
\definecolor{lightblue}{RGB}{121,124,210}
\definecolor{lightpurple}{RGB}{153,102,255}
\definecolor{red}{RGB}{178,34,34}
\definecolor{gray}{RGB}{166,166,166}

\newcommandx{\guest}[3][1=]
    {\setulcolor{lightorange}{\ul{#1}} \textcolor{lightorange} 
    {[\textbf{#2:} #3]}}
\newcommandx{\jane}[2][1=] 
    {\setulcolor{lightgreen}{\ul{#1}} \textcolor{lightgreen}   
    {[\textbf{Jane:} #2]}}
\newcommandx{\ryan}[2][1=] 
    {\setulcolor{lightblue}{\ul{#1}} \textcolor{lightblue}  
    {[\textbf{Ryan:} #2]}}
\newcommandx{\jeff}[2][1=] 
    {\setulcolor{lightpurple}{\ul{#1}} \textcolor{lightpurple}  
    {[\textbf{Jeff:} #2]}}
\newcommandx{\luke}[2][1=] 
    {\setulcolor{lightred}{\ul{#1}} \textcolor{lightred}  
    {[\textbf{Luke:} #2]}}
    
\newcommand{\mone}{\textbf{M1: Ubiquitous}}
\newcommand{\mtwo}{\textbf{M2: Discoverable}}
\newcommand{\mthree}{\textbf{M3: Contextual}}
\newcommand{\mfour}{\textbf{M4: Recoverable}}

\newcommand{\tapIcon}[1]{
    \includegraphics[height=#1]{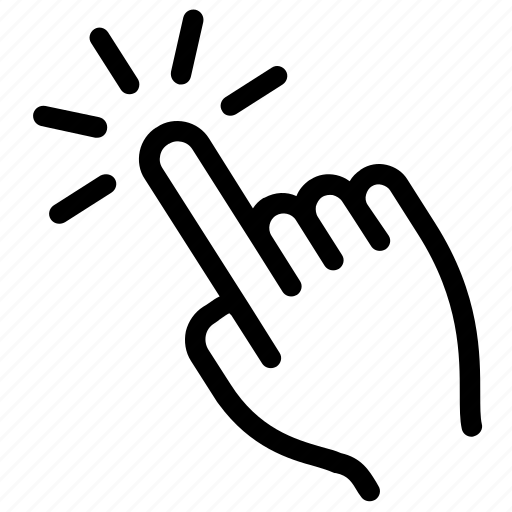}
}

\newcommand{\dragIcon}[1]{
    \includegraphics[height=#1]{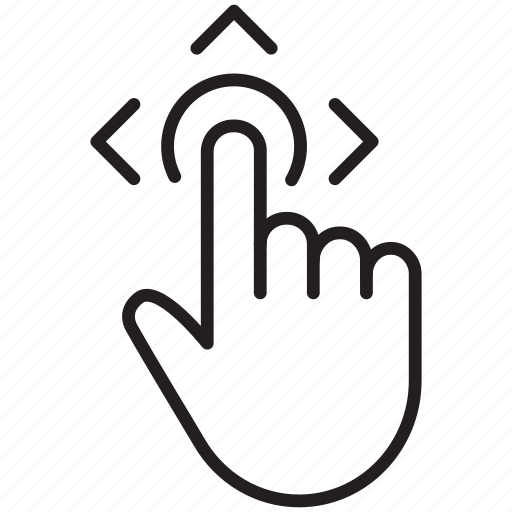}
}

\newcommand{\motionIcon}[1]{
    \includegraphics[height=#1]{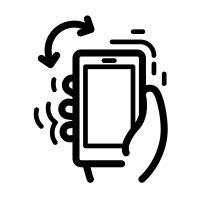}
}

\newcommand{\menuIcon}[1]{
    \includegraphics[height=#1]{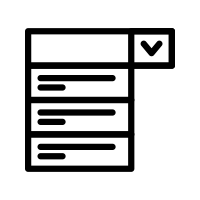}
}

\begin{document}

\teaser{
    \centering
    \includegraphics[width=1\textwidth]{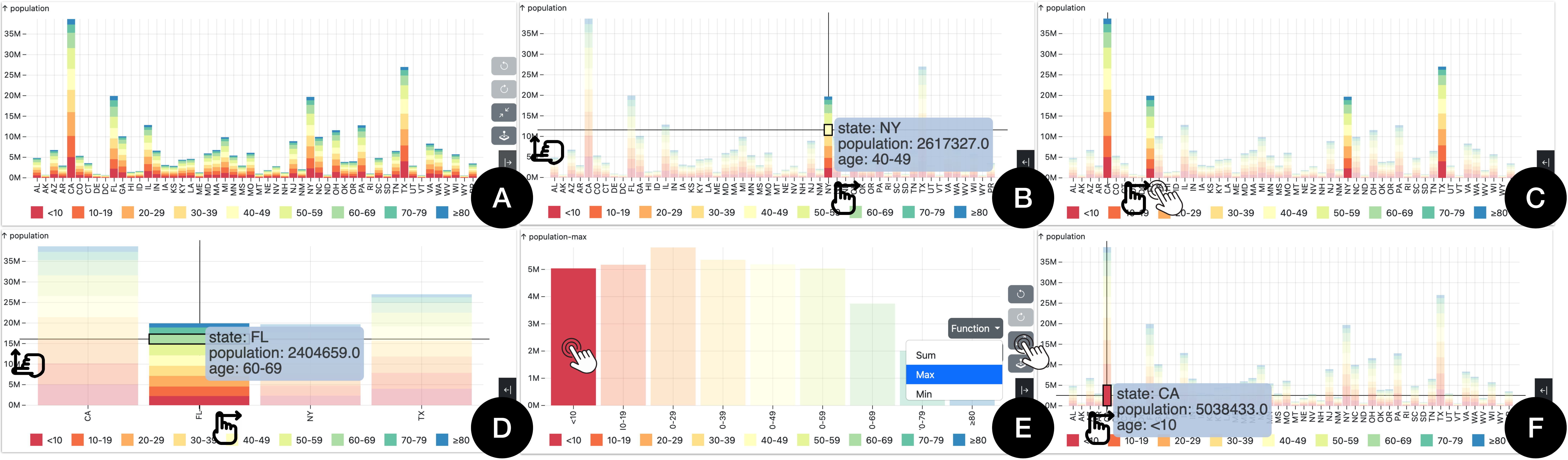}
    \caption{Interactions for mobile exploratory data visualization: (A)~a stacked bar chart of US population data by age, with the menu opened on the right; 
    (B)~the user drags along both axes to \textsc{inspect} the data via a tooltip; 
    (C)~the user drags along the $x$-axis to \textsc{inspect} and taps to \textsc{select} the states with the highest population (CA, FL, NY, TX); 
    (D)~to aid comparison, the user double taps to \textsc{focus} on the selected marks and drags along both axes to further \textsc{inspect} them; 
    (E)~to return to the initial view from Step 1, the user shakes their device to \textsc{reset}, and taps the \textsc{aggregate} button to show the \texttt{MAX(population)} for each \texttt{age} group; the user then taps the $\texttt{age} < 10$ mark; 
    (F)~to \textsc{inspect} the selected mark in the original context, the user taps the \textsc{aggregate} button again to return to the unaggregated view, which shows that CA has the largest population for $\texttt{age} < 10$. A demo video showcasing these interactions is available online at \url{https://osf.io/e2ng8/}.
    }
    \label{fig:teaser}
}

\maketitle

\begin{abstract}
The ubiquity and on-the-go availability of mobile devices makes them central to many tasks such as interpersonal communication and media consumption.
However, despite the potential of mobile devices for on-demand exploratory data visualization, existing mobile interactions are difficult, often using highly custom interactions, complex gestures, or multi-modal input.
We synthesize limitations from the literature and outline four motivating principles for improved mobile interaction: leverage ubiquitous modalities, prioritize discoverability, enable rapid in-context data exploration, and promote graceful recovery.
We then contribute thirteen interaction candidates and conduct a formative study with twelve participants who experienced our interactions in a testbed prototype.
Based on these interviews, we discuss design considerations and tradeoffs from four main themes: precise and rapid inspection, focused navigation, single-touch and fixed orientation interaction, and judicious use of motion.

\end{abstract} 
 

\section{Introduction}
\noindent Nine-in-ten Americans own a smartphone~\cite{pewresearch}, making mobile devices a key form factor for on-the-go tasks, including navigation, shopping, interpersonal communication, and media consumption.
Exploratory data visualization exhibits similar potential for mobile utility, such as on-demand health and fitness tracking or financial monitoring.
However, despite mobile users' desire to engage with interactive, data-driven content~\cite{conlen2019capture}, mobile visualizations are typically static or otherwise retain the desktop interactions in favor of responsive designs that are simpler or more readable~\cite{2020-responsive-visualization}.

Mobile interactions can better facilitate exploratory visualization by retaining the detailed information common for desktop-first designs.
However, current approaches often make use of custom interactions for specific charts~\cite{sadana2014designing, gladisch2015mapping, baur2012touchwave, schwab2019evaluating}, complex gestures~\cite{isenberg2012gestures, wobbrock2009user, rzeszotarski2014kinetica, xu2022enabling}, or multi-modal input such as voice and pen~\cite{srinivasan2020inchorus, jo2017touchpivot}.
These interactions can be inconsistent across different applications or use cases, making them difficult to discover and remember.
Mobile's limited touch vocabulary (tap and drag combinations) results in a tension between interactions that are simple and easy to learn while being expressive enough for users to disambiguate their intents~\cite{lee2021mobile}.

We synthesize the key limitations from prior work to motivate four design principles for improved mobile interaction: (1)~leverage ubiquitous modalities to provide consistent access on different devices and applications; (2)~prioritize discoverability to ensure ease-of-use; (3) enable rapid, in-context data exploration to support mobile efficiency; and (4)~promote graceful recovery for immediate error correction.
We then propose thirteen mobile interactions rooted in basic touch and motion.
We show how these interactions can apply to three common visualizations (scatter, bar, multi-line) through interviews with twelve participants using a testbed prototype. 
Our results highlight design considerations to facilitate (1)~precise and rapid data inspection, (2)~focused navigation, (3)~single-touch and fixed orientation options, and (4)~judicious use of motion.

\section{Related Work}
Extensive prior work has explored different interaction paradigms, such as pen interfaces~\cite{brehmer2013multi, jo2017touchpivot}, gestures~\cite{isenberg2012gestures, wobbrock2009user, xu2022enabling}, tactile stimulation~\cite{wang2019augmenting}, motion and spatial interaction~\cite{kister2017grasp, langner2017v, besanccon2021state}, augmented reality~\cite{goh20193d}, and voice or natural language input~\cite{srinivasan2020inchorus, kim2021data}.
Multi-modal systems~\cite{srinivasan2020inchorus, jo2017touchpivot} can provide flexibility for interactive visualizations, allowing users to disambiguate their intents across preferred interactions; however, their practical adoption for mobile remains limited.
For instance, pen input may not always be available and has typically been used for larger displays such as tablets~\cite{srinivasan2020inchorus, jo2017touchpivot}, whereas voice input may not be suitable in many social contexts.
We ground our interactions in touch and motion to ensure that they can be used anywhere and anytime.

While people have historically expressed a preference for touch inputs over traditional WIMP (Windows, Icons, Menus, Pointer) interfaces~\cite{drucker2013touchviz}, current touch interfaces are often specialized to specific chart types, such as time series~\cite{brehmer2018visualizing, schwab2019evaluating}, stock charts~\cite{apple}, scatter plots~\cite{sadana2014designing}, stacked graphs~\cite{baur2012touchwave}, and networks~\cite{gladisch2015mapping, eichmann2020orchard}.
Such idiosyncratic designs typically lack discoverability or fail to generalize to other visualization types.
For instance, Brehmer~\etal~\cite{brehmer2018visualizing} implement touch-based selection over time ranges, but this interaction breaks down when marks overlap or occlude each other, as in a typical scatter plot.

Lee~\etal~\cite{lee2021mobile} outline other post-WIMP interactions for mobile devices (e.g., accelerometers, haptic feedback, GPS, and cameras). Lee~\etal further note core issues with existing mobile interactions that motivate our work, such as ``fat-fingering'' (i.e., unintended touch) and occlusion due to limited mobile screen space, as well as a limited touch vocabulary, which is exhibited by many current tools, e.g., Vega-Lite~\cite{satyanarayan2016vega}, Observable Plot~\cite{observable}, Mosaic~\cite{heer2023mosaic}, and DIVI~\cite{snyder2023divi}. These tools either limit the expressiveness of possible interactions, or necessitate more complex gestures that are hard to discover and  remember~\cite{wobbrock2009user}.
To our knowledge, this work is the first to contribute a consistent and expressive set of mobile interactions that utilize simple, direct inputs (i.e., touch and motion) to generalize across visualization types.

\begin{table*}[!h]
\centering
\setlength\tabcolsep{0pt} 
\smallskip 
\def\arraystretch{0.70}
\scalebox{1}{
    \begin{tabular*}{\linewidth}{@{\extracolsep{\fill}} cllc}
        \toprule
        \textbf{Task} & \textbf{Mechanism} & \textbf{Intent} & \textbf{Input(s)} \\
        \toprule
            \multirow{2}{*}{Inspect} & Drag fingers along axes (two-finger inspection) & \multirow{2}{*}{Highlight mark for details-on-demand} & \multirow{2}{*}{\dragIcon{5mm}} \\
                                     & Drag finger in $x/y$ direction (single-finger inspection) & & \\
            \midrule 
            \multirow{3}{*}{Select} & Drag w/ lasso & Select marks within lasso region & \multirow{3}{*}{\tapIcon{5mm}\dragIcon{5mm}} \\
                                    & Tap mark & Select mark or legend group & \\
                                    & Tap axis & Select actively inspected mark(s) & \\
                                    
            \midrule
            Focus & Double tap & Focus (zoom + inclusive filter) selection & \multirow{2}{*}{\tapIcon{5mm}\dragIcon{5mm}} \\
            Remove & Quickly swipe & Remove selection from view (exclusive filter) & \\
            \midrule
            \multirow{3}{*}{Aggregate} & Select merge from menu & Aggregate active selection (defaults by $x$-axis encoding) & \multirow{3}{*}{\menuIcon{6.5mm}} \\
                                       & Select encoding from menu & Aggregate selection by encoding & \\
                                       & Select aggregate operator from menu & Change aggregation function & \\
            \midrule
            Reset & Quickly shake or tilt & Reset view & \multirow{2}{*}{\motionIcon{7mm}\menuIcon{6.5mm}} \\
            Undo & Select undo from menu & Undo interaction view & \\
            Redo & Select redo from menu & Redo interaction view & \\
            \bottomrule
        \end{tabular*}
    }

    \vspace{-2pt}
    \caption{Our thirteen interaction candidates. Each candidate describes the \textsc{task} that the user wants to perform, the \textsc{mechanism} to execute the interaction, the user \textsc{intent}, and the \textsc{input} type from touch (direct tap\tapIcon{2.5mm}, drag\dragIcon{2.5mm}, or menu\menuIcon{3mm}) and motion (shake or tilt\motionIcon{3mm}) modalities.}
    \vspace{-5.5mm}
    \label{table:interaction-candidates}
\end{table*}
\section{Mobile Interactions}
\label{section:design-space}
To address the lack of consistent, easy-to-use mobile interactions for exploratory data visualization, we contribute thirteen interactions rooted in simple touch and motion (Table~\ref{table:interaction-candidates}).
In this section, we present four motivating principles that inform our proposed interaction candidates. 
We then implement these interactions in a visualization system developed as a platform to test and explore these interactions with users (\S\ref{section:study}).
Our interactions are best illustrated in our video demo available online at \url{https://osf.io/e2ng8/}.

\subsection{Motivating Principles}
We surveyed prior art, starting from Lee~\etal's~\cite{lee2021mobile} survey of interaction for mobile visualization, to identify existing limitations and four overarching principles for improved mobile interaction:

\bpstart{M1: Leverage ubiquitous modalities.} 
Mobile interactions should use modalities that can be leveraged anywhere and anytime, avoiding those that may be unavailable (e.g., pen~\cite{srinivasan2020inchorus, jo2017touchpivot}) or inappropriate in certain contexts (e.g., voice~\cite{srinivasan2020inchorus}).
Modalities that apply only to a given context would require implementation of multiple modalities for the same task, and result in a larger interaction space that might be more difficult for users to remember.

\bpstart{M2: Prioritize discoverability.} 
Mobile interactions should utilize simple, familiar gestures (e.g., tap, swipe, pinch, or spread). Complex designs for specific tasks or visualizations~\cite{brehmer2018visualizing, schwab2019evaluating, sadana2014designing, baur2012touchwave, gladisch2015mapping, eichmann2020orchard, apple} offer increased expressivity at the expense of discoverability and ease-of-use. 

\bpstart{M3: Enable rapid, in-context data exploration.} 
On-demand mobile visualization necessitates \textit{efficient} exploration, including the ability to quickly inspect and select data while maintaining context for comparison.
Existing interactions often fail to do so, in part due to limited screen space, mark occlusion, and ``fat-fingering'' (e.g., unintended touch inputs when selecting a small mark)~\cite{lee2021mobile}.
Users are generally forced to tediously select individual marks of interest, navigating to reduce clutter as needed (e.g., zooming in to select overlapping points and then zooming out for context).

\bpstart{M4: Promote graceful recovery.} 
Mobile interactions should make discrete changes to a visualization, allowing users to quickly return to a previous state if desired. 
Stateful interactions are vital for efficient correction on mobile devices given that unintended touch actions are more common with small screens~\cite{lee2021mobile}.

\subsection{Candidates}
\label{section:interaction-candidates}
Guided by our design principles, we derived a set of thirteen interaction candidates (Table~\ref{table:interaction-candidates}), which were refined from discussions during our formative study (\S\ref{section:interview-results}).
These candidates use touch and motion (\mone{}) with basic inputs (\mtwo{}), including tap, double tap, drag, and shake / tilt. 
Our thirteen interaction candidates cover six common visualization tasks~\cite{brehmer2013multi}:

\vspace{6px}\bpstart{\textsc{inspect}:}
Users can drag their fingers along an axis to insert a vertical or horizontal line (Fig.~\ref{fig:teaser}B) that displays a tooltip for intersecting marks to aid rapid exploration (\mthree{}).
By using two fingers (one for each axis), the user can quickly select a single point to inspect.
One finger can also be used for dual-axis inspection; to aid discoverability, we add a joystick button (Fig.~\ref{fig:teaser}A) to enable one-handed mode (\mtwo{}).
This interaction was carefully designed to resolve mobile issues with ``fat-finger'' selection due to limited screen space~\cite{lee2021mobile} (\eg needing to continually zoom in and out in order to select the correct point).
Based on feedback from our formative study participants~(\S\ref{section:interview-results}), we also implemented even-spaced control when inspecting marks.
Specifically, for a given axis inspection line, we count the number of intersected marks and divide the inspection range ($x$-axis \texttt{width} or $y$-axis \texttt{height} for two-finger interaction, or \texttt{thumb range} for single-finger interaction) for even, step-wise movement.
This change resolves difficulty with inspecting neighboring marks, exhibited by popular tools that use closest distance~\cite{observable, heer2023mosaic}.

\bpstart{\textsc{select}:}
We retain familiar selection interactions for better discoverability (\mtwo{}): tap a mark~(Fig.~\ref{fig:teaser}C, E), tap a legend mark to select the group, or drag to select an area.

\bpstart{\textsc{focus}:}
Users can quickly navigate to an active selection via double tap (Fig.~\ref{fig:teaser}D), which removes unselected marks and rescales the axes (\ie zooms) to fit the selection.
\textsc{focus} stores the prior view's state, allowing users to quickly return via \textsc{undo} (\mfour{}) if they wish to explore another area (\mthree{}).
Our initial \textsc{focus} interaction decoupled filter and navigation, but participants from the formative study~(\S\ref{section:interview-results}) noted that this coupling would be faster (\mthree{}) and, when combined with \textsc{inspect}, forgoes the need to continually zoom in and out.

\bpstart{\textsc{remove}:}
Users can \textit{exclusively} filter an active selection (\ie remove the selected points, rather than focusing on them) by quickly swiping the screen, as suggested by prior art~\cite{drucker2013touchviz}.

\bpstart{\textsc{aggregate}:}
Users can aggregate an active selection, change the attribute being aggregated by, and modify the aggregation function via menu options (Fig.~\ref{fig:teaser}E).
Temporal and quantitative aggregation automatically bins the data, providing useful on-demand distribution metrics (\mthree{}).
We used direct manipulation for \textsc{aggregate} initially, such as tapping an axis to aggregate by the corresponding attribute, but participants in our formative study preferred menu options to aid discoverability (\mtwo{}).

\bpstart{\textsc{reset / undo / redo}:}
Users can reset the visualization by shaking or tilting their mobile device energetically, as well as return to a previous state (\mfour{}) via undo / redo icons~(Fig.~\ref{fig:teaser}E) for faster recovery and exploration (\mthree{}).
Our initial prototype used motion for \textsc{undo / redo}, but frustrated some participants~(\S\ref{section:interview-results}) who found the frequent use of motion tiresome.
\vspace{-2.1px}
\section{Formative Study}
\label{section:study}
We conducted a formative study with twelve participants to assess the usability of our proposed interaction candidates;
we then leveraged these results to further iterate on our interactions (\S\ref{section:interaction-candidates}).

\vspace{-2.1px}
\subsection{Methods}
We conducted semi-structured interviews with twelve participants: one software engineer, ten graduate students, and one undergraduate, all from computer science.
Each interview was conducted over Microsoft Teams and lasted about one hour.
Participants joined via both desktop and phone to better record how they \textit{physically} interacted with their mobile devices. For example, we observed if they interacted in a one- or two-handed fashion, what orientation they gravitated towards, and how they used motion (e.g., shaking or tilting).
The interviews began with an open-ended discussion of issues faced when exploring data on mobile devices.
Participants then performed each interaction candidate across several visualizations. 
Finally, participants shared their overall impressions and feedback.

Participants interacted with three common chart types: (1)~a scatter plot of the Iris dataset~\cite{irisdataset}, (2)~a bar chart showing United States population data~\cite{uspopulationdataset}, and (3)~a~multi-line chart of unemployment data from the Bureau of Labor Statistics~\cite{blsdataset}.
We asked participants to perform each interaction individually (e.g., ``\textsc{inspect} any mark'', ``\textsc{Remove} any mark'', and ``\textsc{aggregate} any marks'') to ensure coverage over all interactions.
After performing each candidate, participants were asked for immediate impressions and open-ended feedback on the ease-of-use (\mone{}, \mthree{}), discoverability (\mtwo{}), and suggested improvements.
We did not ask open-ended analysis questions, though such questions may be useful to assess learnability in future work~(\S\ref{section:discussion}).

\subsection{Results}
\label{section:interview-results}
The first author performed open coding on the qualitative feedback. 
We then identified the following four primary categories:

\noindent\textbf{Precise and Rapid Inspection.}
During the discussion of existing challenges with mobile interactions, all participants described issues with selecting data due to the limited screen space and ``fat-fingering.''
One participant noted that interleaving text and visual information, as in the case of news content, sometimes leads to unintended selection of nearby text (P3).
This limitation often encouraged participants to prioritize desktop devices for data exploration and analysis to ensure precise, controlled exploration via mouse hovering (an interaction often overlooked on mobile). In fact, two participants reflected on this mismatch between available mobile interactions, and common analysis tasks: \myquote{I have a smart watch and so I have the app on my phone where I can look at...health data. There have definitely been moments where...I expected to be able to hover} (P6);
\myquote{Inspect is what happens when I hover over something with my mouse and select is what happens when I click. But in a touch [device] you don't have the hover equivalent} (P10).

When using our prototype, participants reacted enthusiastically to our \textsc{inspect} interaction: \myquote{This is super cool though... It's like so satisfying even to just do this}~(P1); \myquote{Honestly, I see this being very useful. I really like the control}~(P5); \myquote{I've never seen this joystick convention before and I do like it. I think it's a really novel solution to the inspect problem on touch}~(P10).
Based on feedback from P6, we implemented even-spaced control in lieu of nearest-mark selection (c.f., Observable Plot~\cite{observable}) to prevent flickering when inspecting neighboring marks: \myquote{it's like pretty finicky because like a slight movement of my hand will no longer [inspect] the circle.}

\noindent\textbf{Inspect and Select, then Navigate.}
Participants indicated frustration with mobile zooming during the initial discussion, \eg \myquote{Sometimes you zoom in and it accidentally interacts with what you're trying to zoom into} (P2).
Limited screen space also results in frequently zooming in to select occluded marks and then zooming out for context (e.g., to compare points), which can be tedious:
\myquote{If you just keep zooming in and out then that...becomes annoying} (P8).
After the formative study, we added our \textsc{focus} interaction for users to quickly navigate to selected marks for more detailed analysis, replacing the need to continually zoom in and out.
Based on P7 and P8's suggestions, \textsc{focus} filters and zooms to an active selection via double tap: \myquote{Could be this zoom goes one step ahead in that it's drawing a bounding box and cutting the points outside that box}~(P7); 
\myquote{you tap on something and then it... zooms in}~(P8).

\noindent\textbf{Handedness \& Orientation Preferences.}
Participants' everyday use of their mobile device influenced much of their reactions to our prototype. 
Many participants (9/12) stated that they liked our prototype's use of familiar modalities (\mone{}) and direct manipulation gestures (\mtwo{}).
Most participants preferred single-finger interactions that could be performed without needing to change the way they hold their device: \myquote{I like the simpler ones that require one finger because...if you have to coordinate two fingers, that's...less accessible in general} (P4).
Many participants also preferred interactions that could be performed entirely in a single orientation (e.g., not needing to switch between portrait and landscape for different interactions).
When asked to switch to landscape orientation for the bar and multi-line charts, 11 out of 12 participants needed to access their phone settings to disable locked orientation.
Reliance on landscape orientation has caused frustration in the past, with P3 observing that \myquote{Landscape mode is given more attention but not as convenient.}

However, some participants (3/12) preferred two-handed interactions that can be performed in landscape mode.
P8 liked the precision of the two-finger version, and P10 remarked that \myquote{If you're doing exploring, I think you're a bit more invested and you're going to hold your phone sideways and you're gonna get both fingers in there.}
Our original \textsc{inspect} interaction required two fingers to move along both axes. 
However, given preference for one-handed support, we also developed a single-finger version to control movement along both axes, equivalent to a joystick (as P10 noted).
Participants who used this newly implemented, single-finger version highlighted \myquote{that it's easier because the thumb is far away from the actual point} (P7).
One participant was left-handed, but did not experience any issues since single and dual-axis interactions (e.g., $y$-axis inspection) can be performed on either side of the screen.


\noindent\textbf{Judicious Use of Motion.}
Some participants (3/12) disliked that existing mobile interactions could not be quickly undone without refreshing.
This limitation can discourage interaction and reinforce people's preference for static content.
Our prototype initially used motion input to undo the most recent interaction, but participants experienced usability issues due to unintended movement:
\textit{``shaking is just something that happens unintentionally, especially if I'm walking with my phone''} (P4). P2 also quipped, \textit{``I don't use motion a lot to control things other than like Mario Kart.''}
We thus decided to use motion only for \textsc{reset} to immediately restart exploration, and implemented a higher acceleration threshold to accommodate users' baseline movement. 
Reactions were positive after this update: \textit{``Oh, I just cleaned the graph. Nice''} (P5).
We then added menu options for \textsc{undo} and \textsc{redo} (\mfour{})~(\S\ref{section:interaction-candidates}).
Future work should continue to explore new quality of life improvements, such as the ability to lock/unlock these motion interactions.
\section{Limitations \& Future Work}
\label{section:discussion}
While encouraged by the results from our initial formative study, continued testing and refinement with more users is needed to support general adoption.
We imagine desktop will often be preferred, especially for more targeted analysis, but expect that for our common, exploratory use cases, mobile will provide a more direct and lightweight option.
This separation may also delineate an interesting space of ``hand-off'' transitions between mobile and desktop interaction.
We also plan to test semi-automated interactions that may expedite exploration, such as generalized selection~\cite{heer2008generalized}.

Many of the existing challenges of mobile interaction are likely to be exacerbated by other form factors, like smartwatches, although we believe our principles (\eg~\mtwo{}) and design themes (\eg~judicious use of motion) can still inform future work.
For instance, smaller devices might benefit from thoughtfully integrating our techniques with recent responsive visualization work~\cite{2023-dupo, 2022-cicero-responsive-grammar, 2020-responsive-visualization}, employing interaction when needed for more detailed information.
While participants liked our use of simple modalities and gestures, further evaluation is needed to assess \textit{learnability}, which could be tested by measuring the time it takes users to remember our interactions weeks after initial use.

\bibliographystyle{eg-alpha-doi} 
\bibliography{bibliography}

\end{document}